\documentstyle[12pt]{article}
\begin{document}
\title{Quantum Mechanical Operator of Time}
\author{Slobodan Prvanovi\'c \\
{\it Institute of Physics, University of Belgrade, P.O. Box 57, } \\
{\it 11080 Belgrade, Serbia }}
\date{}
\maketitle

\begin{abstract}
The self adjoint operator of time in non-relativistic quantum
mechanics is found within the approach where the ordinary
Hamiltonian is not taken to be conjugate to time. The operator
version of the reexpressed Liouville equation with the total
Hamiltonian, consisting of the part that is a conventional
function of coordinate and momentum and the part that is conjugate
to time, is considered. The von Neumann equation with quantized
time is found and discussed from the point of view of exact time
measurement.
\\ PACS number(s): 03.65.Ca, 03.65.Ta
\end{abstract}

\section{Introduction}

Time in quantum mechanics (QM), as is well known, appears as a
c-number parameter, not as an operator representing observable. In
this way QM differs from special relativity where time and space
coordinates are treated on an equal footing. Absence of time
operator in QM is strongly related to the finite lower bound of
the energy spectrum. Namely, according to Pauli, it is not
possible to find self adjoint operator which is canonically
conjugate to Hamiltonian for it has bounded-from-below spectra.
Since this objection, there have been many attempts to address
time and/or its operator in QM. The extensive lists of references
one can find in [1-2], while the present approach have some
concrete similarities with those given in [3-7].

Here, operator of time $\hat t$ is found together with its
conjugate one $\hat s$ and these two emerged after their classical
mechanics (CM) counterparts ($t$ and $s$) have been analyzed. As
would become obvious, appearance of $\hat t$ and $\hat s$ is
consistent with the standard QM and $\hat t$ is as was expected -
self adjoint and with continuous spectra.

This article is organized as follows. In Sec. II. discussion of of
the position of time in CM is given. In Sec. III. the self adjoint
time operator, new form of dynamical equation of QM and
possibility of exact measurement of time in QM are analyzed in
details. Finally, in Sec. IV. some remarks are given.

\section{Time and Dynamical Equation of Classical Mechanics}

If time and space coordinates should be on an equal footing, then
there should be operator of time $\hat t$ which has similar
commutation relation with its conjugate operator, say $\hat s$, as
coordinate $\hat q $ has with $\hat p $. ({\sl Nota bene}, it is
not guaranteed {\sl a priori} that the Hamiltonian is conjugate to
time, so that is the reason for taking $\hat s$.) If the Lie
bracket of QM, which is ${1\over i \hbar}[ \ , \ ]$, of $\hat t$
and $\hat s$ does not vanish, then the Poisson bracket, being the
Lie bracket of CM, should not vanish for $t$ and $s$, the last
being the CM counterpart of $\hat s$. This is necessary since
there should be 1-1 correspondence between QM and CM. On the other
hand, perhaps the most important place in CM where Poisson bracket
appears is the Liouville equation:
\begin{equation}
{\partial \rho \over \partial t }= \{ H, \rho \}=
{\partial H\over \partial q }\cdot {\partial \rho \over \partial p } - {\partial H \over
\partial  p }\cdot {\partial \rho \over \partial q }.
\end{equation}
In this expression derivation with respect to $s$ is not
(manifestly) present, so (it seams that) time and coordinate are
not on an equal footing since there are ${\partial \over \partial
p}$ beside ${\partial \over \partial q}$. However, if ${\partial H
\over
\partial s} = 1$, ${\partial H\over \partial t} = 0$ and $\rho
=\rho (q,p,t,s)$, then the LHS of (1) is equal to:
\begin{equation}
{\partial H\over \partial s }\cdot {\partial \rho \over \partial t } - {\partial H\over
\partial  t }\cdot {\partial \rho \over \partial s }.
\end{equation}
If (2) is taken instead of the LHS of (1), then the complete
Liouville equation is:
$$
{\partial H\over \partial s }\cdot {\partial \rho \over \partial t } - {\partial H\over
\partial t }\cdot {\partial \rho \over \partial s }=
$$
\begin{equation}
={\partial H\over \partial q }\cdot {\partial \rho \over \partial p } - {\partial H\over
\partial  p }\cdot {\partial \rho \over \partial q }.
\end{equation}
Now, one can introduce:
\begin{equation}
\{  , \} _{q,p} ={\partial \over \partial q }\cdot {\partial \over \partial p } - {\partial \over
\partial  p }\cdot {\partial \over \partial q },
\end{equation}
\begin{equation}
\{ , \} _{t,s} = -{\partial \over \partial t }\cdot {\partial \over \partial s } + {\partial \over
\partial  s }\cdot {\partial \over \partial t },
\end{equation}
and
\begin{equation}
{\bf \{ } , {\bf \} }_{W} =\{  , \} _{q,p} - \{  , \} _{t,s} ,
\end{equation}
which act on ordered pair $(A(q,p,t,s),B(q,p,t,s))$. Obviously, in
(6) coordinate and time are on an equal footing and one can
reexpress (3) as:
\begin{equation}
{\bf \{ } H, \rho {\bf \} }_{W} = 0.
\end{equation}
So, if ${\partial H \over \partial s}=1$ coordinate and time can
be equally treated in the generalized Liouville equation (7).
Therefore, it is necessary to deduce the relation between $H$ and
s. The first choice is $H=s$, the meaning of which is that
Hamilton function is the conjugate variable to time in CM. But,
this would imply that the Hamiltonian is the conjugate observable
to time in QM. Assuming this, one would find oneself faced with
the problem of time in QM, which was mentioned above. Therefore,
$H$ is not equal to $s$, but:
\begin{equation}
H=H(q,p)+s.
\end{equation}
With this form of Hamilton function one should proceed in
quantization in order to avoid collision with well known facts of
QM. But, before addressing quantization of the above proposed
formalism, few comments are in order. Due to (8), $s$ is present
in (7), but it is ineffective in sense that solutions of this
dynamical equation are:
\begin{equation}
\rho (q,p,t,s)=\rho (q,p,t) \delta (s-s_o ).
\end{equation}
$\delta (s-s_o )$ in (9) ensures that $\rho (q,p,t,s)$ is pure
state in case when $\rho (q,p,t)=\delta (q-q(t)) \delta (p-p(t))$.
If for $t=t_a$ one calculates the mean value of $H(q,p)$ according
to:
\begin{equation}
\int \int \int H(q,p) \rho (q,p,t_a ,s) dq \ dp \ ds ,
\end{equation}
then it would be the same as the mean value of $H(q,p)$ calculated
for $\rho (q,p,t_a )$ in standard phase space formalism of CM,
where $s$ is not considered. On the other hand, if one calculates
the mean value of $H(q,p) + s$ for (9), then it would differ from
(10) in additional $s_o$. However, the measurement of energy is
formalized by the application of $H(q,p)$, not but by $H(q,p)+s$,
so one can take $s_o = 0$ since this value is irrelevant for
physics.

\section{Time and Dynamical Equation of Quantum Mechanics}

Regarding QM, there should be $\hat t$ which has continuous
spectra. Beside this operator, there should be $\hat s$ (and $\hat
q$ and $\hat p$). Since:
$$
\{ t , s\} _{t,s} = -1,
$$
it should be:
\begin{equation}
{1\over i\hbar } [ \hat t , \hat s ] = -1.
\end{equation}
On the other side, existence of $q$, $p$, $t$, $s$, $\{ , \}
_{q,p}$ and $\{ , \} _{t,s}$ resembles the situation within
standard CM when the system with two degrees of freedom is under
consideration (when there are $q_x$, $p_x$, $q_y$ and $p_y$). So,
one can quantize the above given formalism following Dirac and his
procedure appropriate for the case of two degrees of freedom. This
means that one should take ${{\bf\cal H}}_{space} \otimes {{\bf
\cal H}}_{time}$ for the space where operators $\hat q \otimes
\hat I$, $\hat p \otimes \hat I$, $\hat I \otimes \hat t$ and
$\hat I \otimes \hat s$ act. Due to (11), operators $\hat t$ and
$\hat s$ in $\vert t \rangle$ and $\vert s \rangle$
representations should be $t$ and $i\hbar {\partial \over
\partial t}$ and $-i\hbar {\partial \over \partial s}$ and $s$,
respectively. Instead of CM Hamilton function $H=H(q,p) + s$ there
should be its QM counterpart:
\begin{equation}
\hat H = H(\hat q \otimes
\hat I, \hat p \otimes \hat I) + \hat I \otimes \hat s =H(\hat q , \hat p )
\otimes \hat I + \hat I \otimes \hat s .
\end{equation}
Except in the case of free particle, the first part of this
Hamiltonian has, as in standard QM, discrete bounded-from-below
spectrum $E_i$, while the second part has the continuous $s$.

What remains in transition from CM to QM is to declare what is
dynamical equation within this approach. Instead of imitating
procedure related to von Neumann equation in case of two degrees
of freedom, dynamical equation of QM will be reached in another
way.

Without going into details, in [8] it was shown that it is
possible to introduce symmetrized product (ordering rule), denoted
by $\circ$, in standard QM in such a way that:
\begin{equation}
{1\over i\hbar} [\hat H , \hat \rho ] ={\partial \hat H\over \partial \hat q }
\circ {\partial \hat \rho \over \partial \hat p } - {\partial \hat H \over
\partial  \hat p }\circ {\partial \hat \rho \over \partial \hat q },
\end{equation}
where $\hat H = \sum _{m,n} \hat q^m \circ \hat p^n$. Immediate
consequence of (13) is that the von Neumann equation becomes the
operator version of the Liouville equation. Hence, the generalized
Liouville equation (7) is the classical mechanical counterpart of
the operator version of generalized Liouville equation:
\begin{equation}
{\bf \{ } \hat H, \hat \rho {\bf \} }_{W} = 0,
\end{equation}
within which derivations are with respect to the above given
operators, product is the symmetrized one, defined in [8], and
where $\hat H$ is given by (12) while $\hat \rho$ is statistical
operator in ${{\bf\cal H}}_{space} \otimes {{\bf \cal H}}_{time}$.
As was the case for (7), in (14) time and coordinate are treated
equally. From (14) one gets:
\begin{equation}
{\partial \hat \rho \over \partial \hat t}= {1\over i\hbar} [H(\hat q , \hat p ), \hat \rho ],
\end{equation}
where $H(\hat q , \hat p )$ is the first term on the RHS of (12).
The last equation is the von Neumann equation with quantized time.

Due to (11), the equation (15) can be reexpressed:
\begin{equation}
[ \hat s , \hat \rho ]=[ H(\hat q , \hat p ) , \hat \rho ].
\end{equation}
The solutions of (16) are:
$$
\hat \rho = \sum _{i,j} c_{ij} \vert E_i \rangle \langle E_j \vert
\otimes e^{{1\over i\hbar} s_i \hat t} \vert s_o \rangle \langle
s_o \vert e^{-{1\over i\hbar} s_j \hat t}=
$$
\begin{equation}
=\sum _{i,j} c_{ij} \vert E_i \rangle \langle E_j \vert \otimes
\vert s'_i \rangle \langle s'_j \vert ,
\end{equation}
where $H(\hat q , \hat p )\vert E_i \rangle = E_i \vert E_i
\rangle$ and:
\begin{equation}
s_i - s_j = E_i - E_j.
\end{equation}

In case when the system under consideration is in a pure state,
$\hat \rho$ becomes:
\begin{equation}
\vert \psi \rangle \langle \psi \vert = \sum _i c_i \vert E _i \rangle
\otimes \vert s_i \rangle \sum _j c_j ^* \langle E_j \vert \otimes
\langle s_j \vert .
\end{equation}
After substituting this state in (16) and after taking
coordinate-time $\vert q \rangle \otimes \vert t \rangle$
representation, one arrives to:
\begin{equation}
i\hbar {\partial \over \partial t} \psi (q,t) = H(q, -i\hbar
{\partial \over \partial q})\psi (q,t),
\end{equation}
where $\psi (q,t)=(\langle q \vert \otimes \langle t \vert ) \vert
\psi \rangle$. Obviously, (20) is the Schroedinger equation.

Without affecting physical meaning, one can simplify expression
(17) by taking $s_o = 0$ and $s_i = E_i$. Then, in a case when
system is in state with sharp value of energy, $\hat \rho$ becomes
$\vert \psi _i \rangle \langle \psi _i \vert$, where $\vert \psi
_i \rangle = \vert E_i \rangle \otimes \vert E_i \rangle$.
Needless to say, these two $\vert E_i \rangle$'s are pretty much
different. The first one is the element of the Hilbert space
${{\bf\cal H}}_{space}$ and it is normalized to one, while the
second is the element of the rigged Hilbert space ${{\bf \cal
H}}_{time}$, being normalized to $\delta (0)$. When $H(\hat q ,
\hat p)$ acts it "calculates" $E_i$ from $\vert E_i \rangle \in
{{\bf \cal H}}_{space}$, while $\hat s$ just "reads" $E_i$ from
$\vert E_i \rangle \in {{\bf \cal H}}_{time}$. When $\vert E_i
\rangle \in {{\bf \cal H}}_{time}$ is taken in the time
representation $\langle t \vert E_i \rangle$, then $e^{{1\over
i\hbar } E_i t}$ appears. This term emerges in standard QM when
the solution of Schroedinger equation for a system in the
eigenstate of Hamiltonian is discussed. So, it could be said that
the phase factor time dependence of stationary state in standard
QM actually is the time representation of $\vert E_i \rangle \in
{{\bf \cal H}}_{time}$.

Within this approach it holds:
\begin{equation}
[ H(\hat q , \hat p ) \otimes \hat I + \hat I \otimes \hat s ,
\hat I \otimes \hat t ] = i\hbar,
\end{equation}
which means that Heisenberg uncertainty relation for total
Hamiltonian and time holds. However, what one measures is not the
total Hamiltonian, but its "space" part $H(\hat q , \hat p )
\otimes \hat I$, which commutes with time operator $\hat I \otimes
\hat t$. So, one may wonder is it possible to measure energy and
time and to find them with sharp values simultaneously. Assuming
that it is possible, it would mean that the system under
consideration is in state:
\begin{equation}
\vert E_i \rangle \langle E_i \vert \otimes \vert t_a \rangle
\langle t_a \vert ,
\end{equation}
for some $E_i$ and $t_a$. But, states of this form do not satisfy
dynamical equation. Only the states given by (17) are possible
states of physical system and, within their second factor,
eigenstates $\vert s_i \rangle$ of $\hat s$ appear. Since $\langle
s_i \vert t_a \rangle \ne 1$, one can find $t_a$ only with some
probability different from one. Therefore, there would be
dispersion of time for every system that evolves under the action
of some conventional Hamiltonian $H(\hat q , \hat p)$ of QM. In
other words, dynamical equation of QM precludes measurements with
sharp values of time. Measurement of time on a system in
physically meaningful states, which are those satisfying von
Neumann equation, can not deduce exact time. Measurement of
energy, resulting or not in sharp value, is unrelated to this. Of
course, the mean value of $H(\hat q , \hat p)\otimes \hat I$ for a
system in the state (17), calculated according to:
$$
{{\rm Tr}((H(\hat
q , \hat p) \otimes \hat I )\ \hat \rho )\over  {\rm Tr} \hat \rho },
$$
is same as the mean value of this Hamiltonian calculated in
standard QM (where ${{\bf \cal H}}_{time}$ is not taken into
account).

\section{Concluding Remarks}

The variable $s$ was not introduced artificially in CM. Namely,
$t$ has to have conjugate variable, which means that these two
should have non vanishing Poisson bracket. Then, there should be
${\partial \over \partial t}$ and ${\partial \over
\partial s}$ within some new Poisson bracket since the ordinary
Poisson bracket contains derivatives ${\partial \over
\partial q}$ and ${\partial \over \partial p}$ that annihilate
both $t$ and $s$. The Liouville equation, where "space" Poisson
bracket and ${\partial \over \partial t}$ are already present,
lead one not to introduce, but to uncover the exact form of "time"
Poisson bracket.

The intention here was to respect (more strictly than usually) the
request to have time and coordinate on an equal footing. The
generalized Poisson bracket (6), reexpressed Liouville equation
(7) and its operator version (14) are the results of this
intention. On the other hand, (14) offers the possibility to unify
classical and quantum mechanics. Namely, within some generalized
framework one can introduce $h$-dependent ($0\le h \le h_o$)
operators $\hat q _h$, $\hat p _h$, $\hat t _h$ and $\hat s _h$,
all of which become commutative for $h=0$, representing then CM
variables, and noncommutative for $h=h_o$, resembling QM. Within
that framework, both CM and QM will have the same algebraic
structure - the above mentioned symmetrized product, Lie bracket -
operator version of the above given generalized Poisson bracket,
and dynamical equation - the operator version of generalized
Liouville equation. Then, classical and quantum mechanics will
become completely equal regarding their structures and all the
difference will rest on the commutativity of involved operators.
Moreover, that framework will offer the possibility to connect
Newton and Schroedinger equations since they will follow from one
and the same - the operator version of generalized Liouville
equation.

The existence of the self adjoint operator $\hat s$ with
continuous spectrum and its presence in total Hamiltonian is
followed by time-Hamiltonian uncertainty relation. On the other
hand, uncertainty of time is the consequence of dynamics. As the
solutions of Schroedinger equation in standard QM, the solutions
of dynamical equation here time have only within the phase
factors, the consequence of which is its uncertainty. (Present
approach only offers more transparent argumentation on the
impossibility to comprehend exact time.) Of course, this
similarity, as well as the others, does not come as surprise since
one of the major requests of this approach was to keep all
important features of standard QM formalism.

Now, one can approach to the operator of time in reversed order.
Namely, on the LHS of Schroedinger equation (20) one can recognize
$i\hbar {\partial \over \partial t}$ as the time representation of
the operator that is conjugate to time. Then, there should exist
the operator of time, which is $t$ within the same representation.
Finally, one should take ${{\bf \cal H}}_{time}$, beside ${{\bf
\cal H}}_{space}$, where these two operators act since the time
and coordinate within $\psi (q,t)$ are mutually independent. Taken
in this way, it appears that the operator of time was implicitly
present in QM all the time, but it has been just unnoticed.

\section{Acknowledgement}
The work on this article was supported by the Serbian Ministry of
science and technology development through the projects OI 141031
and OI 146012.

\end{document}